\documentclass[pre,aps,twocolumn,showpacs,floatfix,10pt,superscriptaddress]{revtex4}
\usepackage{graphicx}
\usepackage{color}
\usepackage{amsmath, amsthm, amssymb}
\usepackage{epsfig}
\newcommand{\la}{\langle}
\newcommand{\ra}{\rangle}

\newcommand{\be}{\begin{equation}}
\newcommand{\ee}{\end{equation}}	
\newcommand{\bea}{\begin{eqnarray}}
\newcommand{\eea}{\end{eqnarray}}

\begin{document}

\title{Comment on `Self-organized cooperative criticality in coupled complex systems'}
\author{Rahul Dandekar}
\email{dandekar@theory.tifr.res.in}
\affiliation{Department of Theoretical Physics, Tata Institute of Fundamental Research, Homi Bhabha Road, Mumbai 400005, India}







\maketitle

In a recent Letter, Liu and Hu \cite{liuhu} presented a model of toppling-coupled sandpiles, where they found that the avalanche exponents for two toppling-coupled sandpiles are the same as those for a single uncoupled sandpile. In this Comment we provide a proof of this observation for the case when there is conservation of grains in the bulk.\\

Liu and Hu study two sandpiles, denoted by $z_1(i)$ and $z_2(i)$, where $i$ stands for the lattice co-ordinates, which topple when corresponding sites in both piles are above threshold, ie, $z_1(i) \ge z$ and $z_2(i) \ge z$, where $z$ is the co-ordination number of the underlying lattice. In a toplling move, sites topple grains to their respective neighbours ($z_{1,2}(i) \rightarrow z_{1,2}(i)-z$ and $z_{1,2}(j) \rightarrow z_{1,2}(j)+1$ where site $j$ is a neighbour of site $i$). Grains are added independently to both sandpiles, and there is dissipation at the boundaries of the lattice. They find that the probability distribution of avalanche sizes follows a power law with the same exponent as for a single abelian sandpile \cite{btw,ddrev}.\\

Consider the representation of the sandpiles $z_1$ and $z_2$ in terms of a sandpile $z_{min}(i) \equiv \mbox{min}\{z_1(i),z_2(i)\}$, and a process $\Delta z(i) \equiv \lvert z_2(i) - z_1(i) \rvert$ at each site. Also define $z_{max}(i) \equiv \mbox{max}\{z_1(i),z_2(i)\}$. The threshold condition in this representation is simply $z_{min} > 3$ (on a 2D square lattice). When the toppling condition is satisfied, both $z_1$ and $z_2$ topple. Hence, during toppling events, $\Delta z$ does not change. $\Delta z$ changes only when grains are added to either pile: when a grain is added to $z_{min}(i)$, $\Delta z(i)$ decreases by 1 and when grains are added to $z_{max}(i)$ it increases by 1. Addition to $z_1(i)$ results in addition to $z_{max}(i)$ if $z_1(i) \ge z_2(i)$ and $z_{min}(i)$ if $z_1<z_2$. When $z_1(i)=z_2(i)$, that is, $\Delta z(i)=0$, addition to either pile at site $i$ results in addition only to the $z_{max}$ pile.\\

The addition processes, for large times, are independent Poisson processes at each site, and hence the absolute value of their difference, which is the marginal process $\Delta z(i)$, is at each site an independent random walk reflected at the origin. The marginal process $z_{min}$, on the other hand, has the same evolution as a single sandpile on the 2D square lattice, except during addition at a site $i$ where $z_1(i)=z_2(i)$, ie, $\Delta z(i)=0$. Hence, when you consider the sequence of configurations of the sandpile $z_{min}$, sometimes a configuration stays longer than it would in a BTW sandpile process. Thus the observed frequencies of various configurations in $z_{min}$ are different from those in a BTW process by amounts proportional to the probability of finding $\Delta z(i)=0$. Since $\Delta z(i)$ is a random walk reflected at the origin, this probability falls as $t^{-1/2}$, and hence at large times the process $z_{min}$ is exactly the same as a single sandpile, and its steady-state is the same.\\

Because the toppling is controlled by $z_{min}$, so are the avalanches. Thus the avalanche distributions for the coupled sandpiles $z_1$ and $z_2$ are the same as those for the single sandpile $z_{min}$, and hence in the steady-state they are the same as for a single BTW sandpile. A prediction of this argument is that $z_{max}$ averaged over the lattice grows with time as $t^{1/2}$. For a 10x10 lattice, this is shown in figure 1.\\

\begin{figure}[h]
\begin{center}
\includegraphics[scale=0.71]{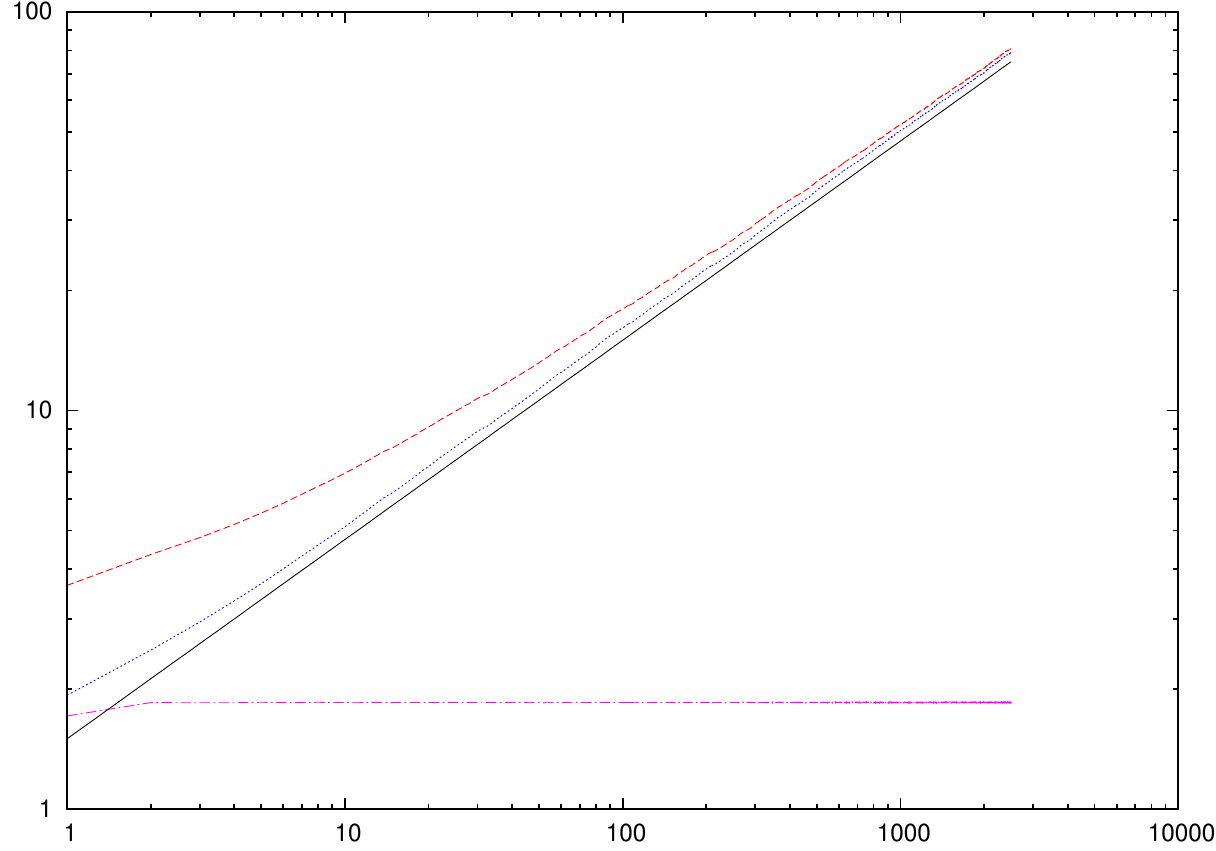}
\end{center}
\caption{$\la z_{max} \ra$ (dashed red line), $\la \Delta z \ra$ (dotted blue line) and $\la z_{min} \ra$ (dot-dashed pink line) for a 10x10 lattice vs time, on a log-log scale. The solid line is the function $1.5~t^{1/2}$.}
\label{move}
\end{figure}

The representation in terms of $z_{min}$ and $z_{max}$ also helps understand the large variations seen in the height profile of either of the sandpiles $z_1$ and $z_2$ in \cite{liuhu}, fig. 5 (b). This is because each site of $z_1$ is randomly either constrained to be less than $4$, or grows as $\sim t^{1/2}$, which results in large uncorrelated variations from site to site.\\

{\bf Acknowledgements -} I thank Vikram Tripathi and Rajdeep Sensharma for discussions. I thank Deepak Dhar for helpful suggestions and a careful reading of the manuscript.

\end{document}